\documentclass[superscriptaddress,nobibnotes,amsmath,amssymb,notitlepage,twocolumn,prl]{revtex4-2}

\usepackage{pdfpages}
\usepackage{etoolbox} 

\usepackage{mathptmx}
\usepackage{braket}
\usepackage{graphicx}
\usepackage[caption=false]{subfig}
\usepackage{hyperref}
\hypersetup{colorlinks=true, linkcolor=blue, citecolor=blue, urlcolor=blue}

\graphicspath{{./img/}}

\newcommand{\beq}[1]{\begin{equation}\label{#1}}
\newcommand{\eep}{\;.\end{equation}}
\newcommand{\eec}{\;,\end{equation}}
\newcommand{\eeq}{\end{equation}}

\renewcommand{\d}{\delta}

\renewcommand{\th}{\theta}

\newcommand{\D}{\Delta}

\DeclareMathAlphabet{\mathcal}{OMS}{cmsy}{m}{n}

\renewcommand{\vec}[1]{{\bf #1}}

\newcommand{\av}{\vec{a}}

\newcommand{\Nz}{N_{\zeta}}

\newcommand{\abinit}{{\sc abinit}~}
\newcommand{\siesta}{{\sc siesta}~}

\begin{document}

\makeatletter
\patchcmd{\@outputpage@head}{\@ifx{\LS@rot\@undefined}{}{\LS@rot}}{}{}{}
\makeatother

\title{Accurate and efficient localized basis sets for two-dimensional materials}


\newcommand{\HarvardPhysics}{Department of Physics, Harvard University, Cambridge, Massachusetts 02138, USA}
\newcommand{\HarvardSeas}{John A.~Paulson School of Engineering and Applied Sciences, Harvard University, Cambridge, Massachusetts 02138, USA}
\newcommand{\Cantabria}{Departamento de Ciencias de la Tierra y Física de la Materia Condensada, Universidad de Cantabria, Avenida de los Castros, s/n, E-39005 Santander, Spain}
\newcommand{\Bath}{Department of Physics, University of Bath, Bath BA2 7AY, United Kingdom}

\author{Daniel Bennett}
\email{dbennett@seas.harvard.edu}
\affiliation{\HarvardSeas}

\author{Michele Pizzochero}
\email{mp2834@bath.ac.uk}
\affiliation{\Bath}
\affiliation{\HarvardSeas}

\author{Javier Junquera}
\affiliation{\Cantabria}

\author{Efthimios Kaxiras}
\affiliation{\HarvardSeas}
\affiliation{\HarvardPhysics}

\begin{abstract}
First-principles density functional theory (DFT) codes which employ a localized basis offer advantages over those which use plane-wave bases, such as better scaling with system size and better suitability to low-dimensional systems.
The trade-off is that care must be taken in order to generate a good localized basis set which is efficient and accurate in a variety of environments.
Here we develop and make freely available optimized local basis sets for two common two-dimensional (2D) materials, graphene and hexagonal boron nitride, for the \siesta DFT code.
Each basis set is benchmarked against the \abinit plane-wave code, using the same pseudopotentials and exchange-correlation functionals.
We find that a significant improvement is obtained by including the $l+2$ polarization orbitals ($4f$) to the basis set, which greatly improves angular flexibility.
The optimized basis sets yield much better agreement with plane-wave calculations for key features of the physical system, including total energy, lattice constant and cohesive energy.
The optimized basis sets also result in a speedup of the calculations with respect to the non-optimized, native choices.
\end{abstract}

\maketitle

\section{Introduction}

\begin{figure*}[t!]
\centering
\includegraphics[width=\linewidth]{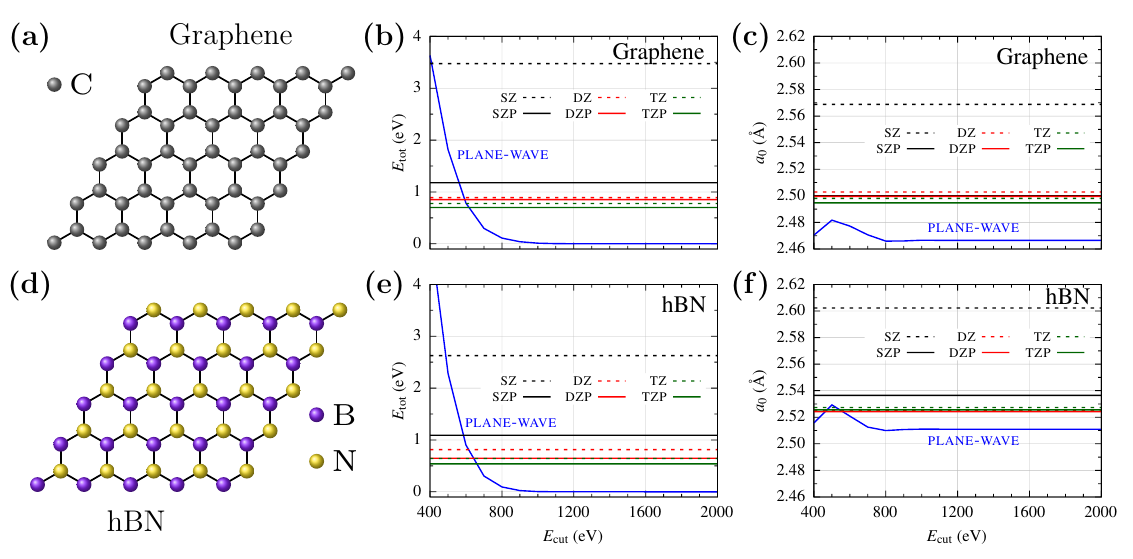}
\caption{
{\bf (a)} Atomic structure of monolayer graphene.
Convergence of 
{\bf (b)} total energy and 
{\bf (c)} lattice constant of monolayer graphene with respect to plane-wave cutoff energy (black), obtained using the \abinit code.
The horizontal lines show the corresponding values obtained from the native local basis sets using the \siesta code: single-zeta (SZ, red), double-zeta (DZ, blue) and triple-zeta (TZ, green), with (solid) and without (dashed) polarization (P) orbitals.
{\bf (d--f)} the same as {\bf (a--c)} for monolayer hexagonal boron nitride.
}
\label{Fig1}
\end{figure*}

Two-dimensional (2D) van der Waals (vdW) materials such as graphene and hexagonal boron nitride (hBN), see Fig.~\ref{Fig1}, are at the heart of much recent interest in condensed matter physics and materials science \cite{geim2007rise}, due to the wealth of physical phenomena they can exhibit, as well as potential applications in nanotechnology.
One crucial recent advance in engineering novel properties in 2D materials is stacking engineering to build stacks of vdW materials with a relative twist angle or strain (lattice mismatch) between layers, referred to as ``twistronics'' or ``straintronics'' \cite{carr2017twistronics}.
These elaborate structures exhibit an interference pattern known as a moir\'e superlattice \cite{Bistritzer2011,andrei2020graphene}.
Devices composed of moir\'e materials display macroscopic quantum phenomena including correlated phases \cite{cao2018correlated,Yazdani2023}, incompressible Hall states \cite{nuckolls2020strongly,wu2021chern} and superconductivity \cite{cao2018unconventional,lu2019superconductors,chen2019signatures,park2021tunable,park2022robust} depending on externally controlled parameters such as the twist angle \cite{inbar2023quantum}, electron density, or electric field.
2D materials also exhibit interesting dielectric \cite{santos2013electrically,laturia2018dielectric} and polar properties, such as interfacial ferroelectricity (which was recently proposed \cite{li2017binary,bennett2022electrically,bennett2022theory} and experimentally observed\cite{yasuda2021stacking,wang2022interfacial,ko2023operando,yasuda2024ultrafast}),
as well as flexoelectricity \cite{springolo2021direct,bennett2021flexoelectric,springolo2023plane}, related with the mechanical coupling between polarization and strain gradients.

Many of these interesting properties, in particular in moir\'e materials, pose a great challenge for computational modeling, as the system sizes are typically very large, from tens to hundreds of nanometers, with supercells containing up to tens of thousands of atoms.
For example, for magic angle graphene, with a twist angle $\th \sim 1.1^{\circ}$ \cite{Bistritzer2011,andrei2020graphene,bennett2024twisted}, the unit cell contains over ten thousand atoms, which is very computationally demanding to simulate using many widely-available electronic structure codes that employ {\it ab initio} methods such as density functional theory (DFT).
Despite the high computational cost, a number of {\it ab initio} studies of moir\'e materials have been performed in recent years, for instance for twisted bilayers of graphene \cite{uchida2014atomic,oshiyama2015large,lucignano2019crucial,cantele2020structural} and transition metal dichalcogenides (TMDs) \cite{naik2018ultraflatbands,vitale2021flat,cheung2024coexisting}.

Another complication is that some of the most widely-used DFT codes, such as \abinit \cite{gonze2009abinit,gonze2016,gonze2020}, {\sc quantum espresso} \cite{giannozzi2009quantum} and {\sc vasp} \cite{kresse1996efficiency}, use a plane-wave basis. 
Although a very high-accuracy can be obtained with these codes because the basis set can be converged systemically, they might not be well-suited for simulating 2D or semi-periodic systems.
The reason for this is that the description of the vacuum region costs the same as the description of the interior of the material, which is an inefficient use of computational resources.

First-principles codes which use a local basis set, such as \siesta \cite{siesta,artacho2008siesta,garcia2020siesta}, {\sc openmx} \cite{ozaki2003variationally,ozaki2004numerical,ozaki2005efficient}, {\sc conquest} \cite{nakata2020large}, {\sc plato} \cite{kenny2009plato}, {\sc adf} \cite{te2001chemistry}, {\sc crystal} \cite{erba2022crystal23}, {\sc cp2k} \cite{kuhne2020cp2k} or {\sc abacus} \cite{chen2010systematically,li2016large}, among others, are more well-suited for 2D layered systems.
Local-basis codes require fewer basis orbitals per atom, which translates into a higher efficiency compared to plane-wave codes, and the vacuum region does not increase the computational cost.
Moreover, using a local-basis set enables the use of linear scaling methods for insulating systems with a well-defined band gap, which makes it possible to simulate very large systems, up to hundreds of thousands of atoms, as implemented in the \siesta \cite{siesta,artacho2008siesta,garcia2020siesta}, {\sc onetep} \cite{skylaris2005introducing} and {\sc conquest} \cite{nakata2020large} codes, for example. 
The tradeoff for this computational advantage is that significant effort is needed in the preparation of unbiased basis sets, in analogy to the extra work required to prepare pseudopotentials to describe the effect of core electrons \cite{junquera2001numerical,anglada2002systematic}. 
In general, a good basis set should be transferable \cite{louwerse2012transferable}: it should be able to describe the electronic degrees of freedom of an atomic species in different environments.
Maximum efficiency is achieved by choosing atomic orbitals that allow convergence with small localization ranges,
but in semi-periodic systems such as 2D materials and surfaces, they should be sufficiently extended in order to describe long-range interactions, and the decay of electron wavefunctions from surfaces into the vacuum \cite{garcia2009optimal,papior2018simple}.

A code like \siesta provides a native basis set with a good tradeoff between efficiency and accuracy, which facilitates the use of the code for routine studies. 
However, depending on the differences in energies between relevant phases in a particular problem, a more refined basis set might be required.
In some cases, the native basis sets generated by \siesta might not result in an accurate description of some material properties, such as total energy and lattice constant, when compared to equivalent well-converged plane-wave calculations, even for simple materials like graphene, see Figs.~\ref{Fig1}(b) and (c).
This limitation is similar to choosing a low energy cutoff in a plane-wave calculation.
The native basis, which is localized on atomic sites, does not sufficiently describe the decay of the electron density into the vacuum.
For the native basis sets, the total energies can be larger by over 1 eV with respect to converged plane-wave calculations, and the difference in lattice constant can be larger than 1\%.

In this work, we develop optimized basis sets for graphene and hBN, which are a significant improvement on the native basis sets generated by \siesta.
In \siesta, higher angular momentum ($l+1$) ``polarization orbitals'' are typically used to improve the angular flexibility of electronic orbitals ($3d$ shell for B,C,N).
We find that the $l+2$ ($4f$ shell) provides much greater angular flexibility due to the three-fold nature of graphene and hBN.
We optimize basis sets with respect to size (number of basis functions for valence electrons), spatial extent (cutoff radii) and angular flexibility (polarization orbitals).
The optimized basis sets give much better agreement with similar plane-wave calculations for several important properties: total energy, lattice constant, electronic bands and cohesive energy.
The basis sets developed in this work are freely available \cite{SM} and will serve as a useful resource for the community which will enable accurate larger-scale simulations involving 2D materials, such as twisted bilayers, multilayers, and stacks of vdW materials.

\begin{figure*}[t!]
\centering
\includegraphics[width=\linewidth]{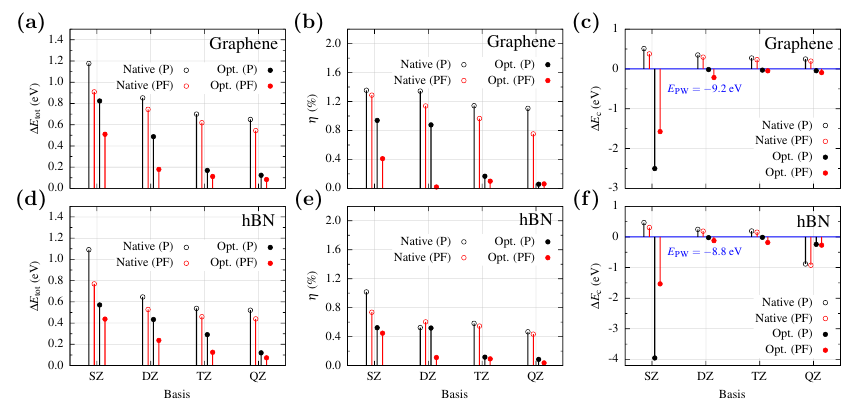}
\caption{
Difference in {\bf (a)} total energy, {\bf (b)} lattice constant (expressed as strain $\eta$) {\bf (c)} cohesive energy of monolayer graphene with respect to \abinit plane-wave calculations using a cutoff of 2000 eV, for basis sets ranging from SZ to QZ, with P (black) and PF (red) polarization orbitals.
The native basis sets are represented by hollow circles and the optimized basis sets are represented by filled circles. 
{\bf (d--f)} the same as {\bf (a--c)} but for monolayer hBN.
}
\label{Fig2}
\end{figure*}

\section{Results}

\subsection{Native basis set}

First-principles DFT calculations were performed using the \siesta \cite{siesta} code, with norm-conserving \cite{norm_conserving} {\sc psml} pseudopotentials \cite{psml}, obtained from Pseudo-Dojo \cite{pseudodojo,footnotepseudo}.
\siesta employs a basis set of numerical atomic orbitals (NAOs) \cite{junquera2001numerical,anglada2002systematic}. 
Calculations were preformed for a series of basis sets of varying size, for both graphene and hBN, where B, C and N have the $2s$ and $2p$ orbitals as valence states.
The number of basis functions $\Nz$, was taken from 1--4. 
Using the notation of \siesta $\Nz =1$ corresponds to single-$\zeta$ (SZ), $\Nz =2$ to double-$\zeta$ (DZ), $\Nz =3$ to triple-$\zeta$ (TZ) and $\Nz =4$ to quadruple-$\zeta$ (QZ).
The ${l+1}$ polarization orbital, where $l$ refers to the maximum angular momentum shell occupied in the free atom (the $3d$ shells for the elements under consideration) was included and denoted by P (SZP, DZP, etc.).
In addition, a set of bases with ${l+2}$ polarization orbitals ($4f$) denoted by F (SZPF, DZPF, etc.), were generated by solving the Schr\"odinger equation for the isolated atoms with the corresponding component of the pseudopotential, as these ${l+2}$ polarization orbitals for the elements under consideration are not included in the native \siesta basis set sizes.

The atomic orbitals in the basis set of \siesta are strictly localized, meaning that they exactly vanish at a given cutoff radius $r_c$, for each of the shells considered.
In order to produce this strict localization on the basis as well as radial functions that are continuous (with all derivatives continuous) at the cutoff radii, a soft confinement potential is added to the atomic hamiltonian used to generate the basis orbitals. 
The functional form proposed in Ref.~\cite{junquera2001numerical} is given by
\beq{eq:soft-confinement}
V(r) = V_0 \frac{\exp \left[- \frac{r_{\rm c} - r_i}{r-r_i} \right]}{r_{\rm c} - r}
\eep
This soft-confinement potential is flat (zero) in the core region, starts off at some internal radius $r_i$ with all derivatives continuous, and diverges at $r_c$ ensuring the strict localization there. 
In the native basis set, the value of the prefactor $V_0$ and the inner radius $r_{\rm i}$ are fixed to 40 Ry and 90\% of the cutoff radii of the corresponding shell, respectively. 
They will be taken as variational parameters when the basis set will be optimized in the following section.

\begin{figure*}[t!]
\centering
\includegraphics[width=\linewidth]{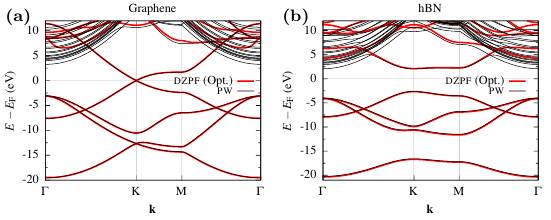}
\caption{
Electronic band structure of monolayer {\bf (a)} graphene and {\bf (b)} hBN, obtained from \abinit calculations (black), using a plane-wave cutoff of 2000 eV, and \siesta (red), using the optimized DZPF basis set.
}
\label{Fig3}
\end{figure*}

A Monkhorst-Pack {$\mathbf{k}$}-point grid \cite{mp} of $20\times 20\times 1$ was used in all calculations, and a fine real space grid was used, determined using a mesh cutoff of 1000 Ry.
The PBE exchange-correlation (XC) functional \cite{pbe} was used in all calculations, from {\sc libxc} \cite{marques2012libxc,lehtola2018recent}.
A vacuum spacing of 40 \AA~was used in all calculations.

Geometry relaxations were performed with each basis set in order to determine the total energy and equilibrium lattice constant. 
The out-of-plane lattice vector (in the direction of the vacuum) was held fixed, and the angle between the in-plane lattice vectors was held fixed at $60^{\circ}$.
The length of the in-plane lattice vectors was relaxed with the constraint ${\left|\av_1\right| = \left|\av_2\right|}$.
The thresholds for the maximum component of the force on any atom and the maximum component of the stress were fixed to 0.1 meV/\AA, and $10^{-4}$ GPa ($6.24\times 10^{-6}$ eV/\AA$^{3}$).
Calculations were also performed using the \abinit \cite{gonze2016,gonze2020} code, which employs a plane-wave basis,
with the same {\sc psml} \cite{psml} pseudopotentials, which are compatible with both codes with the same decomposition into a local-part and non-local Kleinman-Bylander projectors, the same XC functional, and a vacuum spacing of of 40 \AA.
Geometry relaxations were performed in order to calculate the total energy and equilibrium lattice constant as a function of plane-wave energy cutoff, from 400 eV to 2000 eV.

The total energy and lattice constant as a function of basis set size from \siesta and as a function of energy cutoff from \abinit are shown in Fig.~\ref{Fig1}.
The native bases available in \siesta are shown, ranging from SZ(P) to TZ(P).
There is a large disagreement between the codes for both the total energy and lattice constant. 
Even for the largest native basis set (TZP)
\footnote{It is possible to include polarization orbitals with more than one radial function such as~TZDP and TZTP, and so on. 
Here, we only consider single-$\zeta$ for the polarization orbitals.
Increasing the number of basis functions for the polarization orbitals greatly increases the size of the basis set (5 per $\zeta$ for $3d$ and 7 per $\zeta$ for $4f$) and does not yield a significant improvement. 
It is much more efficient to increase the number of radial functions for the valence orbitals.}, 
the total energies differ by over 0.5 eV, and for the smallest basis set (SZ), the energies differ by about 3.5 eV.
The difference in lattice constant (expressed as strain) ranges from 0.5\% to as large as 4\%.

\subsection{Basis set optimization}

Basis sets were optimized using the simplex algorithm \cite{press2007numerical}, following the methodology described in Ref.~\cite{junquera2001numerical}.
During the search of the optimal parameters, the atomic geometry was held fixed. 
In particular, the lattice constants were set to the optimal values from \abinit (2.466 \AA~for graphene and 2.504 \AA~for hBN). 
For these frozen geometries,  the total energy was minimized with respect to several parameters which define the basis set, in particular with respect to
(i) the cutoff radii of the orbitals 
(that control the range of the basis set);
(ii) the prefactor and the inner radii of the soft confinement potentials given in Eq.~(\ref{eq:soft-confinement})  (used to optimize the shape and localization of each angular momentum orbital separately \cite{junquera2001numerical});
and (iii) the net charge of the different atomic species.
Indeed, the shape of an orbital is also sensitive to the ionic character of the atom.
Orbitals in cations tend to shrink, and orbitals in anions tend to expand. 
A parameter $\d Q$ is introduced in the basis for each species, resulting in orbitals better adapted to ionic situations in condensed systems \cite{junquera2001numerical}.
For each atomic species a global $\d Q$, an extra positive or negative charge, is added to the atom at the time of solving the atomic DFT problem for obtaining the basis orbitals.
The extra charge and the confinement potentials are only used to generate the basis, they are not added to the Kohn-Sham Hamiltonian of the system.

Finally, the concept of basis enthalpy is used to tune how localized or extended the basis set is \cite{anglada2002systematic}.
For low-dimensional systems, the basis set must be sufficiently extended such that it describes the decay of the electron wavefunctions into the vacuum correctly.
In order to achieve this, an extra term $p_{\rm basis} V_{\rm orbs}$ was added to the total energy to define an enthalpy, where $V_{\rm orbs}$ is the total volume of the basis set, and $p_{\rm basis}$ is the fictitious ``basis pressure'', which is used as a tuning parameter. 
The minimization of this enthalpy (instead of the total energy) ensures a good tradeoff between accuracy and range of the basis set.
A value of $p_{\rm basis} = 0.03$ GPa was used in order to obtain sufficiently extended bases for 2D materials (the native value is $p_{\rm basis} = 0.2$ GPa). 

Optimized basis sets of were obtained for monolayer graphene and hBN, from SZ to QZ, using P ($3d$) and PF (${3d+4f}$) polarization orbtials.

\begin{figure*}[t!]
\centering
\includegraphics[width=\linewidth]{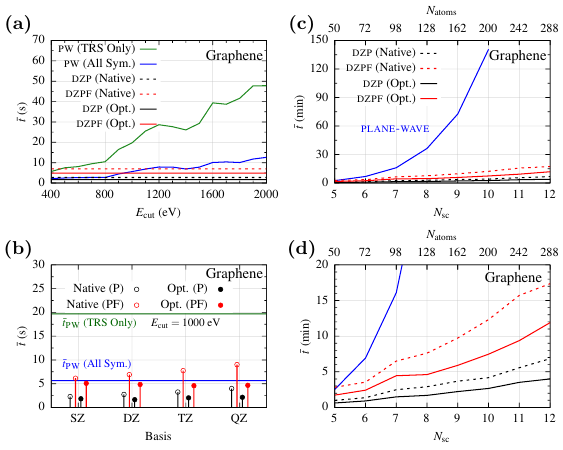}
\caption{
{\bf (a)} Wall time for monolayer graphene with respect to plane-wave (\abinit) calculations as a function of plane-wave cutoff (blue line).
The horizontal lines show the corresponding wall times for \siesta calculations using the DZP (black) and DZPF (red) basis sets, both native (dashed) and optimized (red).
{\bf (b)} Wall time for monolayer graphene for \siesta calculations using native (hollow circles) and optimized (filled circles) basis sets from SZ to QZ, using P (black) and PF (red) polarization orbitals.
The horizontal blue line represents the wall time from an \abinit calculation using a plane-wave cutoff of 1000 eV, for which the total energy and lattice constant are converged.
{\bf (c)}  Wall time per SCF step for $N_{\rm sc} \times N_{\rm sc}$ graphene supercells for $N_{\sc sc}$ ranging from 5 to 12.
Calculations were performed using a single $k$-point ($\Gamma$).
Plane-wave (\abinit) calculations are shown in blue.
{\bf (c)} shows the full range of data and
{\bf (d)} shows a shorter range of times, highlighting the differences between the \siesta calculations.
The number of atoms in each super cell $N_{\rm atoms}$ is indicated above.
Calculations were run on a single core on an isolated node, using as similar parameters as possible for both codes.
}
\label{Fig4}
\end{figure*}

\subsection{Comparison of the optimized basis with plane-waves}

A series of tests were performed in order to compare the optimized basis sets against the native ones, as well as to check which yield better agreement with respect to the \abinit calculations.
Fig.~\ref{Fig2} shows the difference in total energy 
\beq{}
\D E_{\rm tot}=E-E_{\textsc{pw}}
\eec
and strain 
\beq{}
\eta = \frac{a-a_{\textrm{\textsc{pw}}}}{a_{\textrm{\textsc{pw}}}}
\eec
with respect to \abinit at the largest plane-wave cutoff (2000 eV), for native and optimized bases, and with P and PF polarization orbitals.
For basis sets of size DZ and larger, a significant improvement of the total energy and lattice constant is obtained.
Including the $4f$ orbitals almost always improves the agreement of the total energy and lattice constant.

The cohesive energy was calculated as 
\beq{}
E_{\rm c} = E_{\rm mono} - \sum_{i} E^i_{\rm atom}
\eec
namely, the energy of the monolayer minus the energies of the individual atoms in isolation.
The individual atoms were simulated in a cube with a side length of 10 \AA\ for each plane-wave cutoff for the \abinit calculations and for each basis set for the \siesta calculations.
In Fig.~\ref{Fig2} (c) and (f) we show the difference in cohesive energies computed with a plane-wave calculation and with a localized atomic orbital code.
The optimized basis sets give cohesive energies in excellent agreement with plane-wave calculations, excluding SZ, with diminishing returns above DZ.
The optimized SZ basis set gives worse agreement than the native.
A likely reason for this is the basis set superposition error (BSSE).
The basis sets were optimized for monolayer graphene and hBN, meaning that the basis functions for the atoms in one sublattice can improve the description for the atoms in the other sublattice.
This suggests that the SZP and SZPF basis sets are not sufficiently transferable from the condensed systems to the isolated atoms.

Additionally, the optimized basis sets give electronic band structures in excellent agreement with plane-wave calculations, see Fig.~\ref{Fig3}.

\subsection{Timing}

The computational efficiency of the localized basis sets, compared to the native ones and the plane-wave calculations, was also assessed.
The benchmarks were performed using \siesta and \abinit, keeping the calculations as similar as possible.
\abinit version 9.8.4 and \siesta version 5.0.0 were used, both of which were compiled using Intel 2021.8 compilers and the Intel oneAPI Math Kernel Library.
As mentioned previously, both codes used the same {\sc psml} pseudopotentials, and the same XC functionals from the same version of {\sc libxc}.
Each individual calculation was run on a single Intel Cascade Lake core on an isolated node.
For the \abinit calculations, the convergence criteria for the self-consistent field (SCF) loop was a difference in total energy less than $10^{-6}$ Ha.
For the \siesta calculations, the convergence criteria was relative changes of less than $10^{-4}$ for both the density matrix and hamiltonian.
Because the convergence criteria for both codes are not equivalent, the wall time per SCF step was compared: $\bar{t} = t_{\rm wall}/N_{\rm SCF}$, where $t_{\rm wall}$ is the wall time and $N_{\rm SCF}$ is the number of SCF steps.

\siesta only uses time reversal symmetry (TRS) to reduce the number of independent $k$-points.
\abinit uses both TRS and crystal symmetries, which can greatly reduce the number of independent $k$-points for systems with a high degree of symmetry.
\abinit calculations were performed, reducing the $k$-points with TRS only, and with both TRS and crystal symmetries.

Fig.~\ref{Fig4} (a) shows the \abinit wall time per SCF step as a function of plane-wave cutoff energy for graphene using both TRS and crystal symmetries to reduce the $k$-points (blue) and TRS only (green).
The corresponding times from \siesta using DZP and DZPF basis sets, both native and optimized, are represented by the horizontal lines.
Fig.~\ref{Fig4} (b) shows the timings of \siesta calculations using basis sets from SZ to QZ, using P and PF polarization orbitals.
The horizontal lines represent the timings from \abinit with a plane-wave cutoff of 1000 eV, roughly where the calculations are converged.
Adding the $4f$ polarization orbitals significantly increases the computational cost.
Interestingly, the optimized basis sets lead to calculations that are faster than the native ones.

For the smaller simulation box (two atoms per unit cell), when both TRS and crystal symmetries are used in the \abinit calculations, the timings are comparable with the ones obtained with \siesta.
The \siesta calculations are faster than \abinit when the number of $k$-points is reduced using TRS only.
This suggests that the local basis sets are much more efficient for larger systems, where the number of $k$-points is small (or just one), and where the number of atoms within the simulation box increases.
We note that the computation of the hamiltonian and overlap matrix elements in real space are computed with Order-$N$ scaling in \siesta. 
In order to verify this gain in efficiency for larger systems, a series of $\Gamma$-point calculations were performed for $N_{\sc sc}\times N_{\sc sc}$ supercells of monolayer graphene, with $N_{\sc sc}$ ranging from 5 to 12 for \siesta and from 5 to 10 for \abinit.
For the larger system sizes the timing difference between \siesta and \abinit grows very rapidly, as shown in Fig.~\ref{Fig4} (c).
Again, the calculations using optimized basis sets are faster than the calculations using the native basis sets, with the difference between them increasing with system size, see Fig.~\ref{Fig4} (d).
The timing tests for hBN calculations yielded very similar results to those for graphene.

\section{Discussion and Conclusions}

In this work, we investigated the computational efficiency and accuracy of optimized localized basis sets for 2D materials, using graphene and hBN as examples.
We developed and make freely available \cite{SM} optimized basis sets that will serve as a useful tool for simulations of 2D materials using the \siesta code, in particular for calculations involving large supercells.
We generated optimized basis sets from SZ up to QZ, including P and PF polarization orbitals.
We find that the $4f$ ($l+2$) polarization orbitals generally improve agreement with plane-wave calculations.
One possible reason for this is that the $4f$ shells may be more appropriate for graphene and hBN due to their hexagonal nature.
Optimizing the basis sets yielded an excellent improvement in the agreement between energetic and structural properties with respect to converged plane-wave calculations.

Our timing benchmarks demonstrate that localized basis sets can also reduce computational cost.
Calculations using local basis sets offer a speedup compared to comparable calculations using a plane-wave basis.
While \abinit has the ability to further reduce the number of $k$-points with crystal symmetries beyond TRS, significant savings are obtained when the number of $k$-points in the two calculations is equivalent.
In particular, dramatic savings are obtained for $\Gamma$-point calculations using large supercells.
The optimized basis sets yield faster calculations, while still maintaining good agreement for a range of material properties.

In summary, the use of optimized localized basis sets in the \siesta code offers advantages over plane-wave alternatives, particularly in terms of computational efficiency. 
For the 2D materials such as graphene and hBN, the optimized sets developed here provide a balance between computational cost and precision, making them a useful resource for future large-scale calculations.

\section{Acknowledgements}
The computations in this work were run on the FASRC Cannon cluster supported by the FAS Division of Science Research Computing Group at Harvard University.
D.B.~thanks E.~Artacho and M.~J.~Rutter for helpful discussions. 
J.J.~acknowledges financial support from Grant No.~PID2022-139776NB-C63 funded by MCIN/AEI/10.13039/501100011033 and by ERDF ``A way of making Europe'' by the European Union.
D.B., M.P.~and E.K.~acknowledge funding from the US Army Research Office (ARO) MURI project under grant No.~W911NF-21-0147 and from the Simons Foundation award No.~896626.

\clearpage

\includepdf[pages={1}]{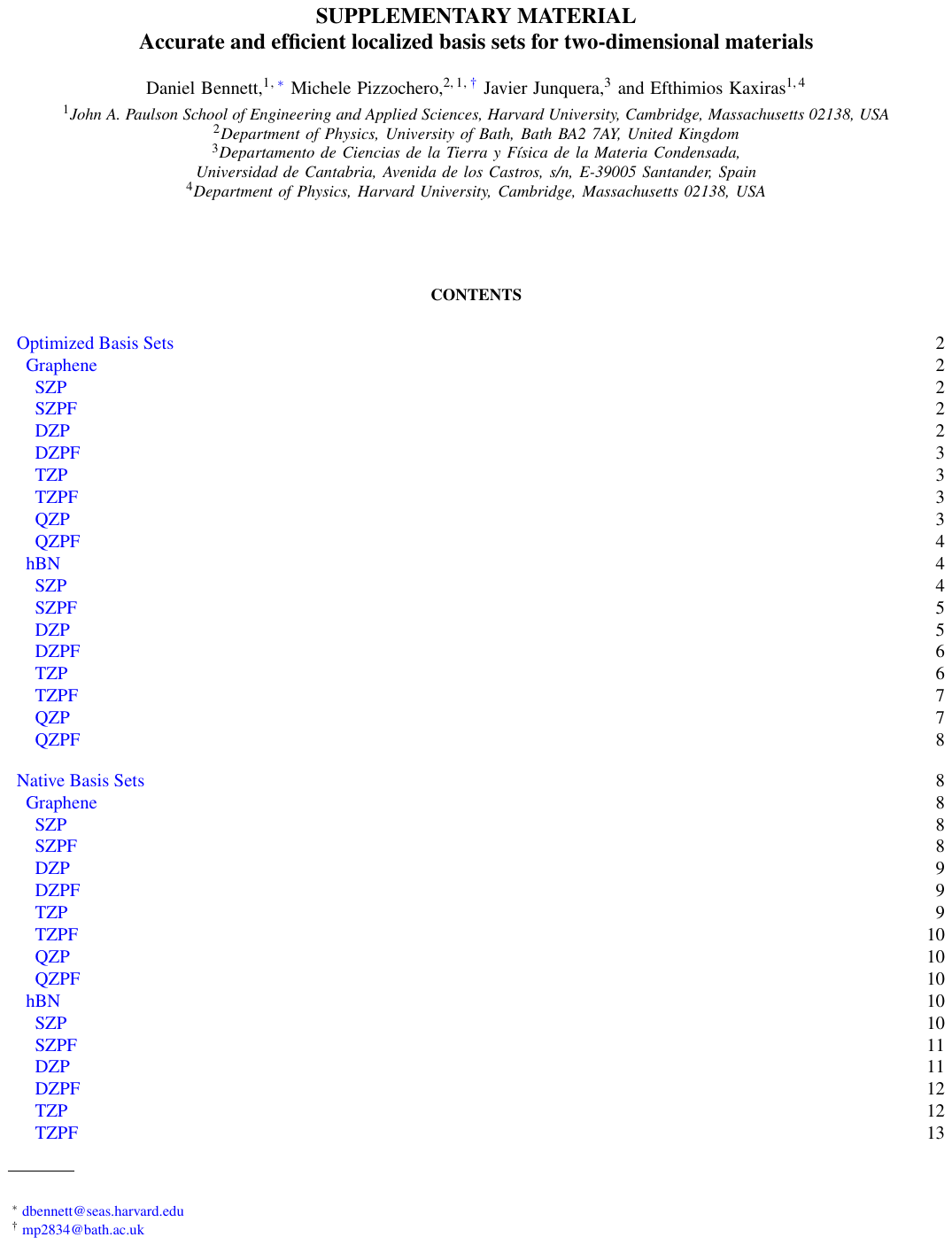}
\clearpage
\includepdf[pages={2}]{./SM.pdf}
\clearpage
\includepdf[pages={3}]{./SM.pdf}
\clearpage
\includepdf[pages={4}]{./SM.pdf}
\clearpage
\includepdf[pages={5}]{./SM.pdf}
\clearpage
\includepdf[pages={6}]{./SM.pdf}
\clearpage
\includepdf[pages={7}]{./SM.pdf}
\clearpage
\includepdf[pages={8}]{./SM.pdf}
\clearpage
\includepdf[pages={9}]{./SM.pdf}
\clearpage
\includepdf[pages={10}]{./SM.pdf}
\clearpage
\includepdf[pages={11}]{./SM.pdf}
\clearpage
\includepdf[pages={12}]{./SM.pdf}
\clearpage
\includepdf[pages={13}]{./SM.pdf}
\clearpage
\includepdf[pages={14}]{./SM.pdf}


\begin{thebibliography}{69}
\bibitem [{foo()}]{footnotepseudo}%
  \BibitemOpen
  \href@noop {} {}\bibinfo {note} {The scalar relativistic \textsc{oncvpsp}
  v0.5 pseudopotentials with standard accuracy were used.}\BibitemShut {Stop}%
\end{thebibliography}
\end{document}